\documentclass[12pt]{article}
\usepackage{cite,epsfig,amssymb,amsmath,graphicx,subfigure,color}\usepackage{caption}
\evensidemargin \oddsidemargin
\evensidemargin \oddsidemargin

\newcommand{\nn}{\nonumber}

\begin{document}

\begin{center}
{{{\Large \bf Holography of Massive M2-brane Theory
\\
with Discrete Torsion}
}\\[17mm]
Dongmin Jang$^{1}$,~~Yoonbai Kim$^{1}$,~~O-Kab Kwon$^{1}$
~~D.~D. Tolla$^{1,2}$\\[3mm]
{\it $^{1}$Department of Physics,~BK21 Physics Research Division,
~Institute of Basic Science, Sungkyunkwan University, Suwon 440-746, South Korea\\
$^{2}$University College,\\
Sungkyunkwan University, Suwon 440-746, South Korea}\\[2mm]
{\it dongmin@skku.edu,~okab@skku.edu,~yoonbai@skku.edu,~ddtolla@skku.edu} }

\end{center}
\vspace{15mm}

\begin{abstract}
We investigate the gauge/gravity duality between the ${\cal N} = 6$ mass-deformed ABJ theory with U$_k(N+l)\times$U$_{-k}(N)$ gauge symmetry and the 11-dimensional supergravity on LLM geometries with SO(2,1)$\times$SO(4)/${\mathbb Z}_k\times$SO(4)/${\mathbb Z}_k$ isometry and the discrete torsion $l$. For chiral primary operators with conformal dimensions $\Delta=1,2$, we obtain the vacuum expectation values using 
the supersymmetric vacua in the mass-deformed ABJ theory and the holographic method in 11-dimensional supergravity. 
We show that the two results agree for all supersymmetric vacuum solutions in the large $N$ limit. The $\frac{l}{\sqrt{N}}$ contributions from the discrete torsion $l$ for  several simple droplet pictures in the large $N$ limit are determined in  holographic vacuum expectation values.  We also explore the effects of the orbifolding ${\mathbb Z}_k$ and the asymptotic discrete torsion $l$, on the gauge/gravity duality dictionary and on the nature of the asymptotic limits of the LLM geometries.
\end{abstract}
\newpage

\section{Introduction}
The duality between the 3-dimensional ${\cal{N}}=6$ Aharony-Bergman-Jefferis-Maldacena (ABJM) theory with U$_k(N)\times$U$_{-k}(N)$ gauge 
symmetry~\cite{Aharony:2008ug} and the M-theory on AdS$_4\times S^7/{\mathbb Z}_k$ geometry, together with the 4-dimensional ${\cal N} = 4$ super Yang-Mills theory, which is dual to type IIB string theory on AdS$_5\times S^5$ geometry~\cite{Maldacena:1997re}, are thoroughly investigated cases of the gauge/gravity duality paradigm~\cite{Maldacena:1997re, Gubser:1998bc,Witten:1998qj}.
The non-conformal extension of the former, which is obtained  by adding an appropriate mass deformation, preserves the full supersymmetry~\cite{Hosomichi:2008jb,Gomis:2008vc}.
The mass-deformed ABJM (mABJM) theory is dual to the M-theory on 11-dimensional Lin-Lunin-Maldacena (LLM) geometries with SO(2,1)$\times$SO(4)/${\mathbb Z}_k\times$SO(4)/${\mathbb Z}_k$ isometry~\cite{Lin:2004nb,Cheon:2011gv}.
Recently, we have made some meaningful tests of this duality~\cite{Jang:2016tbk,Jang:2018aqr,Jang:2017gwd}.
We have calculated the {\it vacuum expectation values} ({\it{vevs}}) of chiral primary operators (CPOs) with conformal dimensions $\Delta=1$ and $\Delta=2$, from the classical vacuum solutions of mABJM theory and from the asymptotic expansion of the LLM solutions using the holographic methods.
In the large $N$ limit, there is an agreement between the {\it vevs} obtained from the two sides.
We  have  also calculated the entanglement entropy by using the Ryu-Takayanagi holographic formula~\cite{Ryu:2006bv} in the asymptotic limit of the LLM geometries and using the path integral methods developed in~\cite{Faulkner:2014jva}.
We have shown that the entanglement entropies obtained by using these two methods agree when the asymptotic limit of the LLM metric satisfies the linearized Einstein equation with an energy-momentum tensor for two scalar fields which encode the information of the mass deformation of the ABJM theory.
See also Refs.~\cite{Kim:2014yca,Kim:2016dzw} for related works.  

An extension of the mABJM theory was achieved by widening the gauge symmetry as U$_k(N+l)\times$U$_{-k}(N)$, where $0\le l<k$~\cite{Aharony:2008gk}.
In the dual gravity theory, the $l$ was identified with the discrete torsion at asymptotic points of the LLM geometry with ${\mathbb Z}_k$ 
orbifold~\cite{Cheon:2011gv}.
In general, the LLM geometries are represented by the droplet pictures, which are infinitely long strips with series of black/white regions along one space direction $\tilde x$~\cite{Lin:2004nb,Cheon:2011gv}.
In these droplet pictures, discrete torsions $\{l_{2i+1},l_{2i}\}$ are assigned to the boundaries between the black/white regions and they count the number of fractional M2-branes at every odd boundary and fractional anti-M2-branes at every even boundary.
In particular, the asymptotic discrete torsion $l$ belongs to the $\tilde x=\pm\infty$ points.  

The structure of the classical vacuum solutions in the extended mass-deformed Aharony-Bergman-Jefferis (mABJ) theory is the same as that of the original mABJM theory, and are expressed in terms of the GRVV matrices~\cite{Gomis:2008vc}.
Evaluation of the $vevs$ from these classical vacuum solutions follows the same line as before.
On the gravity side, we will argue that the LLM geometries with non-vanishing asymptotic discrete torsions are represented by droplet pictures where the origin is shifted by $l$ from the Fermi-level.
As a result, we will show that the discrete torsions satisfy a modified level-matching condition and quantization condition of the 4-form flux. The holographic $vevs$ are expressed as infinite summations involving these modified discrete torsions.
In the general setting, converting these summations into expressions involving the parameters of the theory ($N,k,l$) is out of our reach.
Therefore, we rely on specific examples in order to determine the behaviour of the holographic $vevs$.
In particular, our holographic $vev$ for the CPO of conformal dimension $\Delta=1$ is given by
\begin{align}\label{HRvev0}
\langle {\cal O}^{(\Delta=1)}\rangle_{\rm HR}&=\frac{\mu_0N^{3/2}}{2\pi}
f_0(droplet),
\end{align}
where $\mu_0$ is the mass parameter of the LLM solutions and $f_0(droplet)$ denotes a finite quantity depending on the shape of the droplet only. Interestingly, there is no subleading term in the $\frac{1}{N}$-expansion in the classical limit of the 11-dimensional supergravity.

On the other hand, the result obtained from the classical vacuum solutions {\it on the field theory side} is
given by (see \eqref{vevft})
\begin{align}\label{CLvev0}
\langle {\cal O}^{(\Delta=1)}\rangle_{\rm mABJ}&=\frac{\mu_0N^{3/2}}{2\pi}
\left[f_0(droplet)+\frac {2l}{\sqrt{N}}+\frac{f_1(k,l)}{N}+\cdots\right].
\end{align}
Comparing \eqref{HRvev0} and \eqref{CLvev0}, one can see that in the large $N$ limit, the gauge/gravity duality works exactly for the supersymmetric vacuum solutions of the mABJ theory.
We will show that, in the special case of $k=1$, all the subleading terms are absent and there is an exact agreement between the results obtained from the classical vacuum solutions and from the holographic  methods~\cite{Jang:2016tbk}.
However, in the case of $k>1$, the agreement is achieved only in the large $N$ limit.
The absence of the subleading terms in the $k=1$ case reveals some  meaningful effect of the orbifold singularity on the nature of the gauge/gravity duality dictionary.
 
In order to obtain the holographic $vev$ in \eqref{HRvev0}, we need the asymptotic expansion of the LLM solutions.
The asymptotic limit of LLM geometries represented by any droplet picture is the AdS$_4\times S^7/{\mathbb Z}_k $ geometry.
However, we will show that the radius of the asymptotic $S^7$ is given by
\begin{align}\label{L-rel0}
L=(32\pi^2A_2)^{1/6}l_{\mathrm{P}},
\end{align}
where $l_{\rm P}$ is the Planck length and $A_2$ is determined by the details of the droplet picture of the original LLM geometry.
Therefore, even though every LLM geometry reduces asymptotically to the AdS$_4\times S^7/{\mathbb Z}_k$ geometry, the fact that $L$ depends on $A_2$ suggests that there is some reminiscent information about the original LLM geometries in the asymptotic limit.
 
The remainder of this paper is organized as follows.
In section 2, we discuss the droplet representation of the LLM geometries with the asymptotic discrete torsion $l$ and obtain the appropriate level-matching and quantization conditions.
We also introduce two quantities ($A_2$ and $A_3$), and express them as infinite summations involving the discrete torsions.
These two quantities are crucial to calculate the holographic $vevs$.
In section 3, based on our previous works, we summarize the Kaluza-Klein (KK) reduction of the 11-dimensional gravity on LLM geometries with ${\mathbb Z}_k$ orbifold to obtain 4-dimensional matter-gravity theory on AdS$_4$ background geometry.
In section 4, we evaluate the $vev$ of the CPO of conformal dimension $\Delta=1$ from the classical vacuum solutions of the mABJ theory.
In section 5, we calculate the holographic $vev$ from the asymptotic expansion of the LLM solutions.
We compare the holographic result with classical field theory result.
For few simple droplet pictures  in the large $N$ limit, the $\frac{l}{\sqrt{N}}$ contributions of the asymptotic discrete torsion $l$ are determined in holographic $vev$.
We also comment on the holographic $vev$ for CPO of $\Delta=2$.
In section 6, we draw our conclusions.

\section{LLM Geometry with Discrete Torsion}
Some compelling evidences for the duality between the mABJM theory with U${}_{k}(N)\times$U${}_{-k}(N)$ gauge symmetry and the 11-dimensional supergravity on the LLM geometry with $\mathbb{Z}_{k}$ orbifold, are presented in~\cite{Lin:2004nb, Cheon:2011gv, Jang:2016tbk,Jang:2018aqr}.
It is also conjectured that the mABJ theory with U${}_{k}(N+l)\times$U${}_{-k}(N)$ gauge symmetry is dual to the LLM geometry with $\mathbb{Z}_{k}$ orbifold and asymptotic discrete torsion $l$.
We begin this section by presenting a brief summary of the LLM geometry with asymptotic discrete torsion.

The metric and the corresponding 4-form field strength for the LLM geometry with ${\mathbb Z}_k$ orbifold and SO(2,1)$\times {\rm SO}(4)/{\mathbb Z}_k \times {\rm SO}(4)/{\mathbb Z}_k$ isometry, are given by~\cite{Lin:2004nb}
\begin{align}\label{LLMmetric}
 ds^2 &= -G_{tt} ( -dt^2 + dw_1^2 + dw_2^2) + G_{xx} (d\tilde x^2 + d\tilde y^2) 
+ G_{\theta\theta} ds^2_{S^3/\mathbb{Z}_k}
+ G_{\tilde\theta\tilde\theta} ds^2_{\tilde S^3/\mathbb{Z}_k},
\\
\label{LLMF4}
F_4 &= -d \left(e^{2\Phi}h^{-2}V \right)
\wedge dt\wedge dw_1\wedge dw_2 +\mu_0^{-1} \left[Vd(\tilde y^2e^{2G}) + h^2e^{3G}\star_2 d(\tilde y^2 e^{-2G})\right]
\wedge d\Omega_3
\nn \\
&~~~\,+\mu_0^{-1}
\left[ Vd(\tilde y^2e^{-2G}) -h^2e^{-3G}\star_2 d(\tilde y^2 e^{2G})\right]
\wedge d\tilde\Omega_3,
\end{align}
where $ds^2_{S^3/\mathbb{Z}_k}$ and $ds^2_{\tilde S^3/\mathbb{Z}_k}$ are metrics of two $S^3$'s with $\mathbb{Z}_k$ orbifold while $d\Omega_3$ and $d\tilde\Omega_3$ are the corresponding volume forms~\cite{Auzzi:2009es,Cheon:2011gv}.
The warp factors of the metric and the functions which determine the 4-form field strength $F_{4}$ are given by\footnote{See~\cite{Jang:2016tbk,Jang:2017gwd} for detailed conventions.}
\begin{align}\label{warpfac}
&G_{tt}= -\left(\frac{4\mu_0^2 \tilde y\sqrt{\frac14 - Z^2}}{f^2}
\right)^{2/3},\qquad 
G_{xx}=\left(\frac{f\sqrt{\frac14 - Z^2}}{2 \mu_0\tilde y^2}\right)^{2/3},
\nn \\
&G_{\theta\theta} =\left(
\frac{f\tilde y \sqrt{\frac12 + Z}}{2 \mu_0\left(\frac12 -Z\right)} \right)^{2/3},
\qquad
G_{\tilde\theta\tilde\theta}=\left( \frac{f\tilde y \sqrt{\frac12 - Z}}{2\mu_0 
\left(\frac12 + Z\right)}\right)^{2/3},
\nn\\
&h^2
=
\frac{\sqrt{\frac14 - Z^2}}{\tilde y},
\qquad 
e^{2\Phi}
=
\frac{4 \tilde y \mu_0^2 \sqrt{\frac14 - Z^2}}{f^2},
\qquad 
e^{2 G}
=
\frac{\frac12 + Z}{\frac12 - Z}
\end{align}
with
\begin{align}
f(\tilde x,\tilde y) = \sqrt{1 - 4 Z^2 - 4\tilde y^2 V^2}\,.
\end{align}
The geometry is completely determined by the two functions $Z(\tilde x,\tilde y)$ and $V(\tilde x,\tilde y)$,
\begin{align}\label{ZV}
Z(\tilde x,\tilde y)
=\sum_{i=1}^{2N_b\!+\!1}\frac{(-1)^{i\!+\!1}
(\tilde x\!-\!\tilde x_i)}{2\sqrt{(\tilde x\!-\!\tilde x_i)^2+\tilde y^2}}
\ ,\qquad
V(\tilde x,\tilde y)
=\sum_{i=1}^{2N_b\!+\!1}\frac{(-1)^{i\!+\!1}}{2\sqrt{(\tilde x\!-\!\tilde x_i)^2+\tilde y^2}}.
\end{align}
At $\tilde y=0$, the function $Z(\tilde x,\tilde y)$ has values $\frac12$ for $\tilde x_{2i-1}< \tilde x < \tilde x_{2i}$ and $-\frac12$ for $\tilde x_{2i}< \tilde x < \tilde x_{2i+1}$.
Then the LLM geometries are represented by an infinite strip in the $\tilde x$-direction with regions of $Z(\tilde x,0)=-\frac12$ denoted by black color while regions of $Z(\tilde x,0)=\frac12$ denoted by white color and the $\tilde x_i$'s are the positions of the boundaries between the black and white regions.
See figure 1 below.
The series of black/white regions are called droplets and each diagram contains $N_b$ black/white finite length droplets in addition to a semi-infinite black region below $\tilde x_1$ and a semi-infinite white region above $\tilde x_{2N_b+1}$.

In any droplet picture, if one pulls down all the finite black regions to fill all the finite white regions, then the boundary between the resulting semi-infinite black region and semi-infinite white region is located at
\begin{align}\label{xF}
\tilde x_{\textrm{F}}=\tilde x_1+\sum_{i=1}^{N_b}\tilde b_i,
\end{align}
where $\tilde b_i=\tilde x_{2i+1}-\tilde x_{2i}$ is the length of the $i^{th}$ black region in the original diagram.
This point is referred to as the Fermi level.
In the original diagram, the total length of black regions above the Fermi level is equal to the total length of the white regions below the level.

In Ref.~\cite{Cheon:2011gv}, it was proved that the 4-form flux through the $\mathbb{Z}_{k}$ orbifold of a 4-sphere that surrounds the $i^{th}$ black/white region is proportional to the length of those regions, while the flux through the semi-infinite 4-cycle surrounding the semi-infinite black/white regions is proportional to the total lengths of all finite black/white regions.
As a result, quantization of the fluxes leads to quantization of the lengths of those regions, namely
\begin{align}
\tilde x_{i+1}-\tilde x_i=\frac{2\pi l_{\mathrm{P}}^3\mu_0}{k}\mathbb{Z}.
\end{align}
It was also verified that, introducing a rescaled coordinates ${\rm x}=\frac{\tilde x}{2\pi l_{\mathrm{P}}^3\mu_0}$, the quantized lengths between ${\rm x}_i$ and ${\rm x}_{i+1}$ are related to the discrete torsions of the 3-form gauge field at those two boundaries as
\begin{align}\label{recRel}
l_i+l_{i+1}=\pm({\rm x}_{i+1}-{\rm x}_{i})~{\rm mod}~k,
\end{align}
where the $\pm$ signs are for the black/white regions and $l_i$ is the discrete torsion at ${\rm x}_i$.
The recursion relation \eqref{recRel} is valid modulo $k$ since all the discrete torsions can be put in the range $0\le l_i<k$ by gauge transformation of the 3-form gauge field.

The recursion relation \eqref{recRel} means all the discrete torsions can be fixed once we know the discrete torsions $(l_0=l_{2N_b+2}=l)$ of either at ${\rm x}=\pm\infty$, which are the discrete torsions of the asymptotic geometry AdS${}_4\times S^7/\mathbb{Z}_k$.
It was argued that the discrete torsions at those asymptotic points are related to the rank of the U${}_{k}(N+l)\times$U${}_{-k}(N)$ gauge symmetry of the dual field theory~\cite{Aharony:2008gk, Cheon:2011gv}.
For later convenience, we list the expressions for the discrete torsions at the odd and even boundary points as follows
\begin{align}\label{disTor}
l_{2i+1}&=-\sum_{j=i+1}^{N_b}b_i+\sum_{j=1}^{i}w_i+l~~{\rm mod}~k,\nn\\
l_{2i}&=\sum_{j=i}^{N_b}b_i-\sum_{j=1}^{i}w_i-l~~{\rm mod}~k,
\end{align}
where $b_i={\rm x}_{2i+1}-{\rm x}_{2i}$ and $w_i={\rm x}_{2i}-{\rm x}_{2i-1}$ are the quantized lengths of the $i^{th}$ black and white regions, respectively.

The discrete torsions $l_{2i+1}$ are interpreted as fractional M2-branes at the odd fixed points ${\rm x}_{2i+1}$, whereas the $l_{2i}$ are interpreted as fractional anti-M2-branes at the even fixed points ${\rm x}_{2i}$~\cite{Cheon:2011gv}.
However, it was illustrated that the solutions with anti-M2-branes can be geometrized to yield solutions with M2-branes only.
Those solutions with M2-branes only are compared with vacuum solutions in the dual field theory.
In this case the discrete torsions $l_{2i}$ at the even fixed points are vanishing, i.e., the recursion relation \eqref{recRel} gives
\begin{align}\label{cond1}
{\rm x}_{2i+2}-{\rm x}_{2i}=\mathbb{Z}k.
\end{align}

\begin{minipage}[t]{0.2\textwidth}
%-------------------------------------------------------------------------
         \vspace{-5mm}
         \hspace{-10mm}
\includegraphics[width=1.2\textwidth]{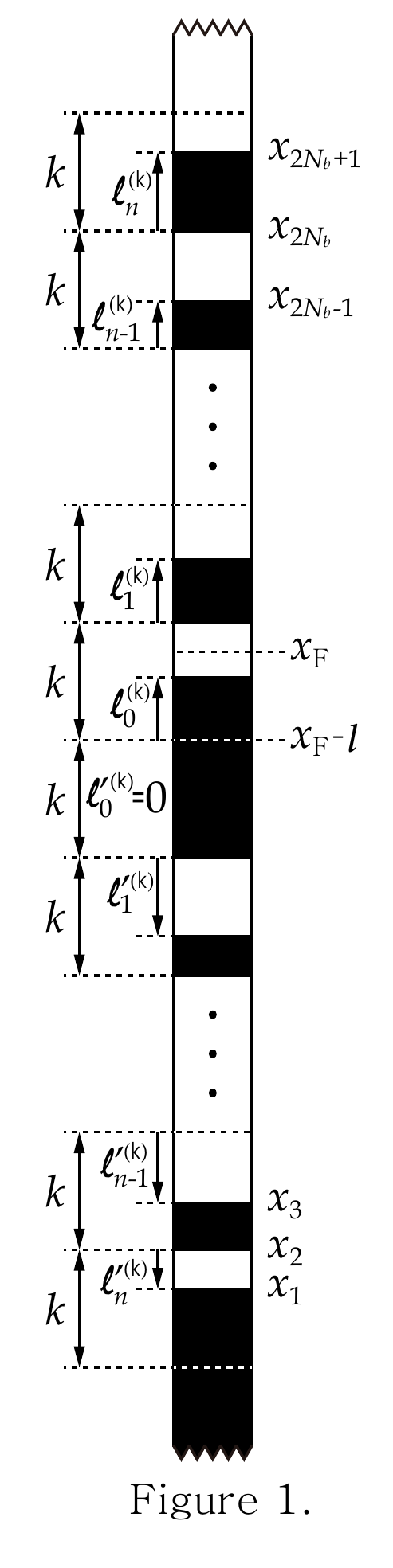}
%-------------------------------------------------------------------------
\end{minipage}
\begin{minipage}[t]{0.75\textwidth}
%\vskip 1.5cm
This means in any droplet picture, the quantized distance between even fixed points is an integer multiple of $k$.
Therefore, we can divide the droplet into intervals of length $k$ and all the even fixed points ${\rm x}_{2i}$ coincide with the boundaries of some of the intervals. See figure 1.

\subsection{Shift of origin in the droplet picture}
Recalling $l_2=0$ and combining \eqref{disTor} and \eqref{xF}, we obtain
\begin{align}\label{cond2}
({\rm x}_{\textrm{F}}-l)-{\rm x}_2=\mathbb{Z}k.
\end{align}
Since ${\rm x}_2$ is at a boundary point of one of the intervals with length $k$, \eqref{cond2} means that the point $({\rm x}_{\textrm{F}}-l)$ is also at one of those boundary points as in figure 1.
For a non-vanishing $l$, we shift the origin of the droplet picture from the Fermi level ${\rm x}_{\textrm{F}}$ to $({\rm x}_{\textrm{F}}-l)$. This shift of origin is required in order to get a correct level-matching condition. See below. The intervals both above and below the shifted origin are labelled by non-negative integers $n=0,1,2,\cdots$ starting at the origin.
Any droplet picture is then parametrized by a set of integers $\{\ell^{(k)}_n,\ell'^{(k)}_n\}$, where $\ell^{(k)}_n$/$\ell'^{(k)}_n$ are the lengths of the black/white region in the $n^{th}$ interval above/below the origin.
These integers are the discrete torsions assigned to the $n^{th}$ interval.
Since the length of the black or white region in a given interval cannot be bigger than the length of the interval itself, these integers should satisfy the condition $0\le \ell^{(k)}_n,\ell'^{(k)}_n\le k$.
\end{minipage}

For the $l=0$ case, counting of the intervals starts at the Fermi level, which implies the level-matching condition $\sum\ell^{(k)}_n=\sum \ell'^{(k)}_n$.
However, in the case of non-vanishing $l$, the counting starts at $({\rm x}_{\textrm{F}}-l)$ point and then this shift of the origin increases the length of the black regions above or/and decreases the white regions below the origin.
As a result, the level-matching condition is now given by
\begin{align}\label{lmc}
\sum_{n=0}^{\infty}\ell^{(k)}_n=\sum_{n=0}^{\infty}\ell'^{(k)}_n+l.
\end{align}
In addition, the relation between the discrete torsions and the number of M2-branes $N$ follows from the quantization of the 4-form flux~\cite{Cheon:2011gv},
\begin{align}\label{NM2}
\sum_{n=0}^{\infty}
\left(
n
+\frac{1}{2}
\right)
\left(
\ell_{n}^{(k)}
+\ell_{n}'^{(k)}
\right)
=
N+\frac{l}{2}.
\end{align}

\subsection{Invariant quantities under the shift of origin}
In the previous subsection, in order to obtain the correct level-matching condition, we have shifted the origin of the droplet picture by $l$ below the Fermi level.
For later convenience, here we discuss two important quantities which are invariant under such shift of the origin.

After some rearranging, the functions $Z(\tilde x,\tilde y)$ and $V(\tilde x,\tilde y)$ in \eqref{ZV} can be written in terms of the Legendre polynomials as follows~\cite{Kim:2016dzw},
\begin{align}\label{ZVfunc}
&Z(r,\xi)=\frac1{2}\Big[\xi+\sum_{n=1}^{\infty}{\rm C}_n\big[(n+1)P_{n+1}(\xi)-2\xi nP_{n}(\xi)+(n-1)P_{n-1}(\xi)\big]
\left(
\frac{2\pi\mu_{0}l_{\rm P}^{3}}{r}
\right)^{n}
\Big],\nn\\
&V(r,\xi)=\frac{1}{2r}\Big[1+\sum_{n=1}^{\infty}{\rm C}_nP_{n}(\xi)
\left(
\frac{2\pi\mu_{0}l_{\rm P}^{3}}{r}
\right)^{n}\Big],
\end{align}
where $\xi =\frac{\tilde x}r$ with 
$r = \sqrt{\tilde x^2+\tilde y^2}$, and $P_n(\xi)$ are the Legendre polynomials. We have also 
introduced~\cite{Jang:2016tbk}\footnote{Note that our definition for ${\rm C}_n$ here differs from that of~\cite{Jang:2016tbk} by factor of $A^{-n/2}$, where $A$ is area of Young diagram.}
\begin{align}\label{Cn}
{\rm C}_n = \sum_{i=1}^{2 N_b +1}(-1)^{i+1}\left(
\frac{{\tilde x_i}}{2\pi\mu_{0}l_{\rm P}^{3}}\right)^n.
\end{align}

The two shift-invariant quantities of our interest are 
\begin{align}\label{A2A3}
A_{2}=\frac12\left({\rm C}_2-{\rm C}_1^2\right),\qquad A_{3}=\frac13 \left({\rm C}_3-3{\rm C}_1{\rm C}_2+2{\rm C}_1^3\right).
\end{align}
These are shift invariants because they can be written in terms of shift invariant variables $b_{i}$ and $w_{i}$, which are the lengths of the black and white regions, as
\begin{align}
A_{2}=\sum_{i=1}^{N_b}\sum_{j=i}^{N_b}w_{i}b_{j},
 \qquad A_{3}=\sum_{i=1}^{N_b}
\left(
\sum_{j=i}^{N_b}\sum_{k=1}^{j}w_{i}b_{j}w_{k}
-\sum_{j=1}^{i}\sum_{k=j}^{N_b}b_{i}w_{j}b_{k}
\right).
\end{align}
These invariant quantities will play central role in the asymptotic expansion of the LLM solutions and consequently in the holographic determination of the vevs of CPOs.

In~\cite{Jang:2016tbk}, we have obtained the expressions of those two invariant quantities in terms of the discrete torsions $\{\ell_n,\ell_n'\}$ for the special case of $(k=1,~l=0)$, where the discrete torsions $\{\ell_n,\ell'_n\}$ are either $0$ or $1$. The results were
 \begin{align}\label{betas}
A_{2}=
\sum_{n=0}^{\infty}\left(n+\frac{1}{2}\right)(\ell_{n}+\ell'_{n}),\qquad A_{3}=\sum_{n=0}^{\infty}n(n+1)(\ell_{n}-\ell'_{n}).
\end{align}
We have also verified that, in this special case, $A_{2}=N$, with $N$ being the rank of the gauge group, and $A_{3}$ is proportional to the vev of a CPO of conformal dimension one in the dual field theory. Here, we follow some indirect procedure in order to obtain the expressions for $A_2$ and $A_3$ for the case of ($k>1$, $0\le l<k$).  To that end, let us rewrite \eqref{betas} by shifting down the origin of the droplet picture by $l$, so that the counting of discrete torsions starts from ${\rm x}_{\textrm{F}}-l$. In order to account for this shift of the origin, we rename the discrete torsions as in the following table
\begin{align}\label{ltilde}
\begin{aligned}
\hline
&
\{\ell_{n},\ell_{n}'\}
,l=0
&
\Longrightarrow
&&&
\{\tilde\ell_{n},\tilde\ell_{n}'\}
,l>0
&&
\textrm{range of }n
\\
\hline
\hline
&
\ell_{n-l}
&
\longrightarrow
&&&
\tilde\ell_{n}
&&
n\geq l
\\
\hline
&
1-\ell_{l-1-n}'
&
\longrightarrow
&&&
\tilde\ell_{n}
&&
0\leq n\leq l-1
\\
\hline
&
\ell_{n+l}'
&
\longrightarrow
&&&
\tilde\ell_{n}'
&&
n\geq 0
\\
\hline
\end{aligned}
\end{align}
In terms of the renamed discrete torsions, the quantities in \eqref{betas} are written as
\begin{align}\label{betas-l}
A_{2}=
\sum_{n=0}^{\infty}\left(n+\frac{1}{2}\right)(\tilde\ell_{n}+\tilde\ell'_{n})-\frac{l^2}2,\quad A_{3}=\sum_{n=0}^{\infty}n(n+1)(\tilde\ell_{n}-\tilde\ell'_{n})-l(l^2-1)-2A_{2}l.
\end{align} 

Next, we write \eqref{betas-l} in terms of the discrete torsions for the case of ($k>1$, $0\le l<k$).
We note that, because of the constraints in \eqref{cond1} and \eqref{cond2}, not all droplet pictures which are allowed in the $k=1$ case are allowed in the $k>1$.
Any droplet picture with arbitrary lengths of the black/white regions is consistent with \eqref{cond1} and \eqref{cond2} in the $k=1$ case, whereas only some subset of those droplet pictures are consistent with those constraints in the $k>1$ case.
For those droplet pictures in the subset, the discrete torsions $\{\ell_n^{(k)},\ell_n'^{(k)}\}$ for $k>1$ can be written in terms of $\{\tilde\ell_n,\tilde\ell_n'\}$ introduced in \eqref{ltilde} as follows
\begin{align}\label{lk}
\ell_{m}^{(k)}=\sum_{n=km}^{k(m+1)-1}\tilde\ell_{n}, \qquad \ell_{m}'^{(k)}=\sum_{n=km}^{k(m+1)-1}\tilde\ell'_{n}.
\end{align}
Rearranging \eqref{betas-l} and using \eqref{lk}, one can rewrite it in terms of $\{\ell_n^{(k)},\ell_n'^{(k)}\}$ as
\begin{align}
&A_{2}
=
k
\left[
\sum_{n=0}^{\infty}
\left(
n
+\frac{1}{2}
\right)
\left(
\ell_{n}^{(k)}
+\ell_{n}'^{(k)}
\right)
-\frac{l}{2}
\right]
-\frac12\sum_{n=0}^{\infty}
\left[
\ell_{n}^{(k)}
\left(
k
-\ell_{n}^{(k)}
\right)
+\ell_{n}'^{(k)}
\left(
k
-\ell_{n}'^{(k)}
\right)
\right]
\nn\\
&
~~~~~~
+\frac12l(k-l),\nn\\
&A_{3}
=
k^{2}\sum_{n=0}^{\infty}n(n+1)
\left(
\ell_{n}^{(k)}
-\ell_{n}'^{(k)}
\right)
-k\sum_{n=0}^{\infty}n
\left[
\ell_{n}^{(k)}
\left(
k
-\ell_{n}^{(k)}
\right)
-\ell_{n}'^{(k)}
\left(
k
-\ell_{n}'^{(k)}
\right)
\right]
\nn\\
&
~~~~~~+\frac13\sum_{n=0}^{\infty}
\left(
\ell_{n}^{(k)3}
-\ell_{n}'^{(k)3}
\right)
-\frac13 l^{3}-2A_{2}l.\label{AV}
\end{align}
These results are crucial in subsequent sections for our holographic calculations of the $vevs$ of CPO with conformal dimensions one and two.

\section{KK Reduction}\label{KK}
In this section, we review the KK reduction of the 11-dimensional supergravity to a 4-dimensional matter-gravity theory, where the matter content of the latter is determined by the LLM solutions.
The KK reduction involves an expansion of the 11-dimensional fluctuations in terms of the spherical harmonics on $S^7/{\mathbb Z}_k$ and then projecting the equations of motions on those spherical harmonics to obtain the equations of motion for various KK modes.
At non-linear order, the resulting equations contain higher derivative terms of the 11-dimensional fluctuations.
Then one needs to introduce the KK maps to absorb the higher derivative terms and obtain the canonical equations of motion for the 4-dimensional modes.
A detailed KK reduction procedure at non-linear order is carried out in~\cite{Jang:2016tbk, Jang:2018aqr,Jang:2017gwd}.
Here, for clarity, we briefly summarize the procedure and write the equations of motion for a few 4-dimensional modes which are necessary for our purpose.

\subsection{Field equations at quadratic order}
In order to write the 11-dimensional gravity equations of motion up to quadratic order in the fluctuations, we perturb the 11-dimensional fields around the AdS$_4\times S^7/{\mathbb Z}_k$ background as
\begin{align}\label{fluct}
{\bf g}_{pq}=g_{pq}+h_{pq}, \qquad {\bf F}_{pqrs}=F_{pqrs}+f_{pqrs},
\end{align}
where $p,q,\cdots=0,\cdots,10$.
The equations of motion for those fluctuations are obtained by inserting \eqref{fluct} into the 11-dimensional gravity equations of motion and keeping all the terms which are quadratic in the fluctuations $h_{pq}$ and $f_{pqrs}$.
The resultant equations are
\begin{align}
&\nabla^r\nabla_{p}h_{qr}+\nabla^r\nabla_{q}h_{pr}-\nabla^2h_{pq}-\nabla_q\nabla_ph^r{}_{r}-Rh_{pq}-g_{pq}\left(-R^{rs}h_{rs}
+\nabla^r\nabla^sh_{rs}-\nabla^2h^r{}_{r}\right)
\nn\\
&\quad\qquad\,\,\,\,\,
+\frac{1}{48}\Big(F_{rstu}F^{rstu} h_{pq}
{-4}g_{pq}h_{rs}F^r{}_{tuv}F^{stuv}\Big)+\frac{1}{24}g_{pq}f_{rstu}F^{rstu}-\frac12 h_{rs}F^r{}_{ptu}F_q{}^{stu}
\nn\\
&\quad\qquad\,\,\,\,\,
-\frac1{6}\Big(
f_{prst}F_q^{~rst}+F_{prst}f_q^{~rst}\Big)=0,
\label{lineq2-1}
\\
\label{C3EoM-1}
&
\nabla_p(h^{t}{}_{t}F^{pqrs})+2\nabla_p(4F_{t}^{~[pqr}h^{s]t}+f^{pqrs})
+\frac2{\sqrt{-g}}\frac{1}{(4!)^2}\tilde\epsilon^{p_1\cdots p_4q_1\cdots q_4qrs}f_{p_1\cdots p_4}F_{q_1\cdots q_4}=0,
\end{align} 
where the indices are raised/lowered by the AdS$_4\times S^7/{\mathbb Z}_k$ background metric and the covariant derivatives as well as the Ricci tensor are also those of the background.

Next we expand the fluctuations in terms of the spherical harmonics on $S^7/{\mathbb Z}_k$, with $ {\rm SO}(4)/{\mathbb Z}_k \times {\rm SO}(4)/{\mathbb Z}_k$ symmetry, so that the orbifolding does not affect the expansion,
\begin{align}\label{metric-exp1}
&h_{\mu\nu}(x,y)=h^{I_1}_{\mu\nu}(x)Y^{I_1}(y),\quad h^\rho{}_{\rho}(x,y)=h^{I_1}(x)Y^{I_1}(y),\nn\\
&h_{(ab)}=s^{I_1}(x)\nabla_{(a}\nabla_{b)}Y^{I_1}(y),\quad h^a_{~a}(x,y)=\phi^{I_1}(x)Y^{I_1}(y),\nn\\
&f_{\mu\nu\rho\sigma}(x,y)=\frac23\nabla_{[\mu} t^{\lambda I_1}(x)\epsilon_{\nu\rho\sigma]\lambda}Y^{I_1}(y),\quad f_{\mu\nu\rho a}(x,y)=- \frac1{3!}\epsilon_{\mu\nu\rho}\!\!~^{\sigma}t_\sigma^{I_1}(x)\nabla_{a}Y^{I_1}(y),\nn\\
&f_{\mu abc}(x,y)=\nabla_\mu t^{I_{35}}(x)Y_{abc}^{I_{35}}(y)
,\quad f_{abcd}(x,y)=4 t^{I_{35}}(x)\nabla_{[a}Y_{bcd]}^{I_{35}}(y),
\end{align}
where $I_{n}=0,1,2,\cdots$.
Here we have split the 11-dimensional indices into the AdS$_4$ indices $(\mu,\nu,\cdots =0,\cdots,3)$ and the $S^7$ indices $(a,b,\cdots=4,\cdots,10)$, $x$ denotes the AdS${}_4$ coordinates, whereas $y$ denotes the $S^7$ coordinates, and ($Y^{I_{1}}$, $Y_{abc}^{I_{35}}$) are the scalar and antisymmetric 3-tensor spherical harmonics on $S^{7}$, respectively.

Plugging \eqref{metric-exp1} into \eqref{lineq2-1} and \eqref{C3EoM-1}, we obtain the equations for various modes. The details can be found in~\cite{Jang:2017gwd}.
Here we copy the equations for the graviton mode $\hat{h}_{\mu\nu}^{0}$ and two scalar modes $\Psi$ and $T$
\begin{align}\label{lineq1-3-06a}
\Big(L_E+\frac{12}{L^2}&\Big)\hat h_{\mu\nu}^0+\frac1{34560}\Big\{-\frac{26}3\nabla_\mu\Psi\nabla_\nu\Psi+\frac{28}3\Psi\nabla_\mu\nabla_\nu\Psi+\frac{L^2}3\nabla_\mu\nabla^\rho\Psi\nabla_\nu\nabla_\rho\Psi\nn\\
&+\frac{L^2}2\nabla^\rho\Psi\nabla_\mu\nabla_\nu\nabla_\rho\Psi+\frac{L^4}{24}\nabla_\mu\nabla^\rho\nabla^\sigma\Psi\nabla_\nu\nabla_\rho\nabla_\sigma\Psi+\frac{L^4}{32}\nabla^\rho\nabla^\sigma\Psi\nabla_\mu\nabla_\nu\nabla_\rho\nabla_\sigma\Psi\nn\\
&-g_{\mu\nu}\Big(\frac{12}{L^2}\Psi^{2}+\nabla^\rho\Psi\nabla_\rho\Psi+\frac{35L^2}{48}\nabla^\rho\nabla^\sigma\Psi\nabla_\rho\nabla_\sigma\Psi-\frac{L^4}{64}\nabla^\tau\nabla^\rho\nabla^\sigma\Psi\nabla_\tau\nabla_\rho\nabla_\sigma\Psi\Big)\Big\}\nn\\
&+\frac{1}{48}\big(T\nabla_{\mu}\nabla_{\nu} T+\frac12\nabla_{\mu} T\nabla_{\nu} T\big)
+\frac{1}{96}g_{\mu\nu}\big(\nabla_\rho T\nabla^\rho T-\frac{16}{L^2} T^{2}\big)=0,
\end{align}
\begin{align}\label{diagpp}
\left(\square+\frac 8{L^2} \right)\Psi+{\cal O}(\mu_0^3)=0,\qquad\left(\square+\frac 8{L^2}\right)T+{\cal O}(\mu_0^3)=0,
\end{align}
where $L_{E}$ represents the Einstein operator, and we have introduced the following diagonal modes
\begin{align}\label{hPsiT}
\hat h^0_{\mu\nu}\equiv h^0_{\mu\nu}-\frac14g_{\mu\nu}\phi^0+\frac1{24}g_{\mu\nu}\hat\psi^0,\qquad \Psi\equiv\frac1{70}(7\hat\psi^2-162\hat\phi^2),\qquad T\equiv t^{I_{35}=1}
\end{align}
with $\hat\phi^{I_1}\equiv \phi^{I_1}+\frac{I_1(I_1+6)}{L^2}s^{I_1},~~\hat\psi^{I_1}\equiv 18 h^{I_1}-L\nabla^\mu t_\mu^{I_1}$.

We use the following field redefinition in order to absorb the higher derivative terms in the graviton equation \eqref{lineq1-3-06a}, and obtain the canonical 4-dimensional graviton mode
\begin{align}\label{FRD1}
H_{\mu\nu}=&\,\hat h^{0}_{\mu\nu}+g_{\mu\nu}\big(C_1\Psi^{2}+C_2\nabla^\rho\Psi\nabla_\rho\Psi\big)+C_3\nabla_\mu\Psi\nabla_\nu\Psi\nn\\
&+g_{\mu\nu}C_4\nabla^\rho\nabla^\sigma\Psi\nabla_\rho\nabla_\sigma\Psi+C_5\nabla_\mu\nabla^\rho\Psi\nabla_\nu\nabla_\rho\Psi+g_{\mu\nu}C_T T^{2},
\end{align}
where
\begin{align}\label{coefs}
& C_1=-\frac{1}{2^6\, 3^3\, 5},\quad C_2=-\frac{L^2}{2^{11}\,3^3\,5},\quad C_3=-\frac{7L^2}{2^{11}\,3^4\,5},\quad C_4=-\frac{L^4}{2^{14}\,3^3\,5},\\
& C_5=-\frac{L^4}{2^{13}\,3^4\,5},\quad C_T=-\frac{1}{2^5\,3}.\nn
\end{align}
Consequently the graviton equation \eqref{lineq1-3-06a} becomes
\begin{align}\label{Hmn-eq4}
\Big(L_E +\frac{12}{L^2}\Big)H_{\mu\nu}&-\frac1{96}\Big( \nabla_\mu T\nabla_\nu T+\frac{M_t^2}2g_{\mu\nu}T^{2}\Big)-\frac1{2304}\Big( \nabla_\mu \Psi\nabla_\nu \Psi+\frac{M_{\psi}^2}2g_{\mu\nu}\Psi^{2}\Big)=0.
\end{align}

\subsection{Asymptotic expansion of LLM geometries}
The $\frac 1r$ expansion of the functions $Z(r,\xi)$ and $V(r,\xi)$ to all orders are given in \eqref{ZVfunc}, however it is difficult to obtain such closed form for the expansions of the functionals in \eqref{warpfac}.
Therefore, for the warp factors in \eqref{warpfac}, we write the asymptotic expansion by keeping only the leading and subleading terms,
\begin{align}\label{astexp}
G_{tt}(r,\xi)
=&\,
-\frac{r^2}{2^{2/3} \pi ^{4/3} l_{\rm P}^4 ({\rm C}_2-{\rm C}_1^2)^{2/3}}
\left(
1
-\frac{2^{2}\pi l_{\rm P}^3(2 {\rm C}_3-3 {\rm C}_2 {\rm C}_1+{\rm C}_1^3)}{3({\rm C}_2-{\rm C}_1^2)r}\mu_{0}\xi
+\cdots
\right),
\nn\\
G_{rr}(r,\xi)
=&\,
\frac{\pi^{2/3} l_{\rm P}^2 ({\rm C}_2-{\rm C}_1^2)^{1/3}}{2^{2/3} r^2}
\left(
1
+\frac{2^{2}\pi l_{\rm P}^3({\rm C}_3-{\rm C}_1^3)}{3({\rm C}_2-{\rm C}_1^2)r}\mu_{0}\xi
+\cdots
\right)
,
\nn\\
G_{\theta\theta}(r,\xi)
=&\,
2^{1/3} \pi^{2/3} l_{\rm P}^2 ({\rm C}_2-{\rm C}_1^2)^{1/3} (1+\xi)
\nn\\
&
\left(
1
+\frac{2\pi l_{\rm P}^3}{3({\rm C}_2-{\rm C}_1^2)r}
\mu_{0}
\left[
(2 {\rm C}_3 -3 {\rm C}_2 {\rm C}_1+{\rm C}_1^3)\xi -3({\rm C}_2 {\rm C}_1-{\rm C}_1^3)
\right]
+\cdots
\right)
,
\nn\\
G_{\tilde\theta\tilde\theta}(r,\xi)
=&\,
-2^{1/3} \pi ^{2/3} l_{\rm P}^2({\rm C}_2-{\rm C}_1^2)^{1/3}(1-\xi)\nn\\
&
\left(
1
-\frac{2\pi l_{\rm P}^3}{3({\rm C}_2-{\rm C}_1^2)r}
\mu_{0}
\left[
(2 {\rm C}_3-3 {\rm C}_2 {\rm C}_1+{\rm C}_1^3)\xi +3( {\rm C}_2 {\rm C}_1 -{\rm C}_1^3)
\right]
+\cdots
\right)
,
\end{align}
where we made the replacement $G_{rr}(r,\xi)\equiv G_{xx}(\tilde x,\tilde y\to r,\xi)$.

By recalling that the asymptotic limit of the LLM geometry is the AdS$_4\times S^7/{\mathbb Z}_k $, the leading terms of the above warp factors should give those of the AdS$_4\times S^7/{\mathbb Z}_k $ geometry. In particular, we should have
\begin{align}\label{lim}
\lim_{r\to\infty}G_{rr}(r,\xi)=\frac{L^2}{4r^2},
\end{align}
where $L$ is the radius of the $S^{7}$.
Comparing \eqref{lim} with the expansion of $G_{rr}(r,\xi)$ in \eqref{astexp}, we obtain
\begin{align}\label{L-rel}
L=(32\pi^2A_2)^{1/6}l_{\mathrm{P}},
\end{align}
where we have used the definition of $A_2$ in \eqref{A2A3}.
With the result in \eqref{L-rel}, we reproduce the AdS$_4\times S^7/{\mathbb Z}_k $ metric by plugging the leading terms of the warp factors in \eqref{astexp} into the LLM metric in \eqref{LLMmetric}.

A comment is in order about the result in \eqref{L-rel}.
In the previous section, we have discussed the LLM geometries which are classified by their droplet representations.
For every droplet picture, we have the corresponding set of the discrete torsions $\{\ell^{(k)}_n,\ell'^{(k)}_n\}$, which are constrained by the level-matching condition \eqref{lmc} and the quantization condition \eqref{NM2}.
The asymptotic limit of all these LLM geometries is the AdS$_4\times S^7/{\mathbb Z}_k $ geometry. However, since the radius $L$ in \eqref{L-rel} depends on $A_2$, which is expressed in terms of the discrete torsions as in \eqref{AV}, there is some reminiscent information about the original LLM geometries in the asymptotic AdS$_4\times S^7/{\mathbb Z}_k $ geometry.

Now using the coordinate transformation
\begin{align}
\rho
=
\frac{L^{3}}{4r},
\end{align}
we rewrite \eqref{astexp} as
\begin{align}\label{astexp2}
G_{tt}(\rho,\xi)
=&\,
\frac{L^2}{2^{2/3}}\frac{1}{4\rho^{2}}
\left(
1
-\frac{2^{4} \pi  l_{\mathrm{P}}^3  (2 C_3-3 C_2 C_1+C_1^3)}{3L^3 (C_2-C_1^2)}\mu _0 \rho \xi
+\cdots
\right),
\nn\\
G_{\rho\rho}(\rho,\xi)
=&\,
\frac{L^2}{4\rho^2}
\left(
1
+\frac{2^{4}  \pi l_{\mathrm{P}}^3 (C_3-C_1^3)}{3 L^3(C_2-C_1^2)} \mu _0 \rho \xi
+\cdots
\right),
\nn\\
G_{\theta\theta}(\rho,\xi)
=&\,
\frac{L^2 }{2}(1+\xi)
\nn\\
&
\left(
1
+\frac{2^{3}\pi l_{\mathrm{P}}^3 }{3 L^3(C_2-C_1^2)}\mu _0 \rho
\left[
(2 C_3-3 C_2 C_1 +C_1^3)\xi-3C_1( C_2-C_1^2)
\right]
+\cdots
\right),
\nn\\
G_{\tilde\theta\tilde\theta}(\rho,\xi)
=&\,
\frac{L^{2}}{2} (1-\xi)
\nn\\
&
\left(
1
+\frac{2^{3}\pi l_{\mathrm{P}}^3 }{3L^3(C_2-C_1^2)}\mu _0 \rho
\left[
(2 C_3-3 C_2 C_1 +C_1^3)\xi+3C_1( C_2-C_1^2)
\right]
+\cdots
\right)
.
\end{align}
In order to read the $vevs$ of gauge invariant operators from the asymptotic expansions of gravity solutions, the Fefferman-Graham (FG) coordinates system is the most convenient.
In the FG coordinates, the LLM metric \eqref{LLMmetric} is written as
\begin{align}\label{dsFG}
ds^2 =&\,\frac{ L^2}{4 z^2}\Big(dz^2 +\frac{4z^2}{L^2} G_{tt}(z,\tau )\left( -dt^2+dw_1^2+dw_2^2\right) \Big)\nn\\
&+ G_{\tau\tau}(z,\tau)d\tau^2+ G_{\theta\theta}(z,\tau ) ds_{S^3/\mathbb{Z}_{k}}^2 + G_{\tilde\theta\tilde\theta}(z,\tau ) ds_{\tilde S^3/\mathbb{Z}_{k}}^2.
\end{align}
The warp factors in \eqref{astexp} are written in the FG coordinates as
\begin{align}\label{warpFG}
&
G_{tt}(z,\tau)
=
\frac{L^2}{4 z^2}-\frac{A_3 L^2 \mu _0 }{2 \sqrt{2} A_2^{3/2} z}\tau
+{\cal O}(\mu_0^2),\nn\\
&
G_{\tau\tau}(z,\tau)
=
\frac{L^2}{4(1- \tau ^2)}+{\cal O}(\mu_0^2),
\nn\\
&
G_{\theta\theta}(z,\tau)
=
\frac{L^2}{2}  (1+\tau)
+\frac{A_3 L^2 \mu _0  z}{2 \sqrt{2} A_2^{3/2}}(1+\tau)^2
+{\cal O}(\mu_0^2),\nn\\
&
G_{\tilde{\theta}\tilde{\theta}}(z,\tau)
=
\frac{L^2 }{2} (1-\tau)
+\frac{ A_3 L^2 \mu _0 z}{2 \sqrt{2} A_2^{3/2}}(1-\tau)^2+{\cal O}(\mu_0^2),
\end{align}
where we have used the following coordinate transformation
\begin{align}
&\rho(z,\tau)
=
z
+\frac{2 (C_1^3-C_3) \mu _0z^2}{3 (C_2-C_1^2)^{3/2}}\tau
+{\cal O}(\mu_0^2),
\nn\\
&\xi(z,\tau)
=
\tau
+\frac{2(C_1^3-C_3) \mu _0 z}{3(C_2-C_1^2)^{3/2}} (\tau ^2-1)
+{\cal O}(\mu_0^2)
.
\end{align}
Similarly, the asymptotic expansions of the 4-form field strength of the LLM solution in \eqref{LLMF4} are given by 
\begin{align}\label{LLMF4exp}
&F_{tw_1w_2z}(z,\tau)
=
-\frac{3 L^3}{8 z^4}
+\frac{3A_{3}L^3 \mu _0 }{2^{5/2}A_{2}^{3/2} z^3}\tau
+{\cal O}(\mu_0^2)
,\nn\\
&F_{tw_1w_2\tau}(z,\tau)
=
-\frac{3A_{3}L^3\mu _0 }{2^{7/2}A_{2}^{3/2} z^2}
+{\cal O}(\mu_0^2)
,\nn\\
&F_{\theta\phi\psi z}(z,\tau)
=
-\frac{1}{8}L^3 \mu _0\sin\theta (1+\tau)^2 
+{\cal O}(\mu_0^2)
,\nn\\
&F_{\theta\phi\psi \tau}(z,\tau)
=
-\frac{1}{4}L^3 \mu _0  z \sin \theta (1+\tau)
+{\cal O}(\mu_0^2)
,\nn\\
&F_{\tilde\theta\tilde\phi\tilde\psi z}(z,\tau)
=
-\frac{1}{8}L^3 \mu _0\sin \tilde{\theta}(1-\tau)^2
+{\cal O}(\mu_0^2)
,\nn\\
&F_{\tilde\theta\tilde\phi\tilde\psi \tau}(z,\tau)
=
\frac{1}{4} L^3 \mu _0 z\sin \tilde{\theta}(1-\tau )
+{\cal O}(\mu_0^2)
.
\end{align}

The expansion in \eqref{warpFG} shows that in the asymptotic region, the warp factors for the LLM metric are given by the AdS$_4\times S^7/\mathbb{Z}_{k}$ background and small fluctuations.
Therefore, in the asymptotic region, the LLM metric are written as in \eqref{fluct}, where the values of the fluctuations $h_{pq}$ are read from the asymptotic expansion in \eqref{warpFG}.
Similarly, from the asymptotic expansion of the 4-form field strength in \eqref{LLMF4exp}, we read the values of the fluctuations $f_{pqrs}$.
Finally, the obtained values of the fluctuations $h_{pq}$ and  $f_{pqrs}$ can be expanded in terms of the spherical harmonics on $S^7/\mathbb{Z}_{k}$ as in \eqref{metric-exp1}, in order to obtain the values of the KK modes ($h^{I_1}_{\mu\nu}, s^{I_1}, \phi^{I_1}, \cdots$).
Here, we are interested only in three physical modes which consist of the graviton mode $H_{\mu\nu}$ and the two scalar modes, $\Psi$ and $T$, introduced in the previous subsection.
From \eqref{hPsiT} and \eqref{FRD1}, we notice that these three physical modes are composed of KK modes ($h^{0}_{\mu\nu},s^{2}, \phi^{2}, \cdots$).
Therefore, we obtain the values of the physical modes by combining the values of the KK modes, which in turn are read from the asymptotic expansion of the LLM geometry.\footnote{Similar procedure was developed in~\cite{Skenderis:2006uy,Skenderis:2006di} to investigate the holographic dual of the Coulomb branch in ${\cal N} = 4$ Yang-Mills theory.}
See Ref.~\cite{Jang:2016tbk} for the detailed procedure.
The results are
\begin{align}\label{HmnTPsi}
&H_{ij} = \left[-\frac{(L\mu_0)^2}{180}\left( 30 + \beta_3^2\right)+{\cal O} \left(\mu_0^4\right)\right]\eta_{ij} , 
\qquad
H_{zz} = - \frac{(L\mu_0)^2}{1440}\left(960 + 29\beta_3^2\right) + {\cal O}\left(\mu_0^4\right),\nn\\
&\Psi = -24\beta_3\mu_0 z+{\cal O}(\mu_0^3),
\qquad T = 16\sqrt{3} \, \mu_0 z + {\cal O}(\mu_0^3),
\end{align}
where $\eta_{ij} = {\rm diag}(-1,1,1)$ and
\begin{align}\label{beta3}
\beta_3 = \frac {3A_3}{A_2^{3/2}}.
\end{align}

\section{Vacuum Solutions of Mass-deformed ABJ Theory and {\it{vevs}} of CPOs}
The mass deformation of the ${\cal{N}}=6$ ABJM theory with $\textrm{U}_{k}(N)\times\textrm{U}_{-k}(N)$ gauge group preserves full supersymmetry but breaks SU(4) global R-symmetry to SU(2)$\times$SU(2)$\times$U(1).
When $l$ overlapping fractional M2-branes are added at $\mathbb{Z}_{k}$ orbifold fixed points, the gauge group of the theory is extended to $\textrm{U}_{k}(N+l)\times\textrm{U}_{-k}(N)$ with $0\leq l<k$.
This theory is often called the ABJ theory~\cite{Aharony:2008gk}.
The vacuum solutions in the mass-deformed ABJ theory are composed of the GRVV matrices, which are also the building blocks of the vacuum solutions in the mass-deformed ABJM theory.

In order to reflect the SU(2)$\times$SU(2)$\times$U(1) global symmetry, the four complex scalars $(Y^{A},~A=1,2,3,4)$ are split into two two-component scalars as
\begin{align}
Y^{A}
=
(Z^{a},W^{\dagger a}),\qquad
a=1,2.
\end{align}
Then, the vacuum solutions are,
\begin{align}\label{vacuum}
&
Z_0^{a}
=
\sqrt{\frac{k\mu}{2\pi}}\oplus_{n=0}^{\infty}
\begin{pmatrix}
\left[M_{(n)}^{a}\right]_{n\times(n+1)}\otimes I_{N_{n}}&0_{n\times n}
\\
0_{n+1\times n+1}&0_{(n+1)\times n}\otimes I_{N_{n}'}
\end{pmatrix}
,
\nn\\
&
W_0^{\dagger a}
=
\sqrt{\frac{k\mu}{2\pi}}\oplus_{n=0}^{\infty}
\begin{pmatrix}
0_{n\times(n+1)}\otimes I_{N_{n}}&0_{n\times n}
\\
0_{n+1\times n+1}&\left[\bar{M}_{(n)}^{a}\right]_{(n+1)\times n}\otimes I_{N_{n}'}
\end{pmatrix}
,
\end{align}
where $\mu=4\mu_{0}$ and $M^a_{(n)}$ are the irreducible GRVV matrices,
\begin{align}
M_{(n)}^{1}
=
\begin{pmatrix}
\sqrt{n}&0&&&&
\\
&\sqrt{n+1}&0&&&
\\
&&\ddots&\ddots&&
\\
&&&\sqrt{2}&0&
\\
&&&&1&0
\end{pmatrix}
&,&
M_{(n)}^{2}
=
\begin{pmatrix}
0&1&&&&
\\
&0&\sqrt{2}&&&
\\
&&\ddots&\ddots&&
\\
&&&0&\sqrt{n+1}&
\\
&&&&0&\sqrt{n}
\end{pmatrix}
\end{align}
with $\bar{M}_{(n)}^{a}=(M_{(n)}^{a})^{\dagger}$. 
Each vacuum solution contains $N_{n}$ irreducible matrices of the first type $M_{(n)}^{a}$ and $N'_{n}$ of the second type $\bar{M}_{(n)}^{a}$.
The set of integers $\{ N_n,N_n'\}$ are called occupation numbers.
Since $(Z^a,W^{\dagger a})$ are $(N+l)\times N$ matrices, the occupation numbers should satisfy the relations,
\begin{align}\label{Nl}
&
\sum_{n=0}^{\infty}
\left(
n
+\frac{1}{2}
\right)
(N_{n}+N_{n}')
=
N
+\frac{l}{2}
,\qquad
\sum_{n=0}^{\infty}(N_{n}-N_{n}')
=
l.
\end{align}
The vacuum solutions are supesymmetric for $0\leq N_{n},N'_{n}\leq k$~\cite{Kim:2010mr}.
There is one-to-one correspondence between the LLM solutions and the vacuum solutions of the mABJM theory~\cite{Cheon:2011gv}.
This correspondence is realized through a one-to-one map between the discrete torsions and the occupation numbers,
\begin{align}\label{NNn}
\{\ell^{(k)}_n,\, \ell'^{(k)}_n\}
\Longleftrightarrow
\{N_n,\, N_n'\}. 
\end{align}

The CPO of conformal dimension one, which is invariant under the SU(2)$\times$SU(2)$\times$U(1) global symmetry with non-vanishing {\it {vev}}, is given by~\cite{Jang:2016tbk}
\begin{align}\label{CPO1}
{\cal{O}}^{(\Delta=1)}
=
\frac14{\rm Tr}
\left(
Z^{a}Z_{a}^{\dagger}
-W^{\dagger a}W_{a}
\right).
\end{align}
In large $N$ limit, the leading term for the {\it{vev}} of this CPO is determined by the classical vacuum solutions in \eqref{vacuum}
\begin{align}\label{vevft}
\langle{\cal{O}}^{(\Delta=1)}\rangle_{\rm mABJ}
&=
\frac14{\rm Tr}
\left(
Z^{a}Z_{a}^{\dagger}
-W^{\dagger a}W_{a}
\right)\Big|_{Z^a,W^{\dagger a}=Z^a_0,W^{\dagger a}_0}+\cdots\nn\\
&=
\frac{k\mu_0}{2\pi}\sum_{n=0}^{\infty}n(n+1)(N_{n}-N_{n}')
+\cdots,
\end{align}
where from here on the ellipses denotes quantum corrections, and $N_{n}$ and $N_{n}'$ satisfy the conditions in \eqref{Nl}.
Using the one-to-one map in \eqref{NNn}, we rewrite the $vev$ in \eqref{vevft} in terms of the discrete torsions of the dual LLM geometry,
\begin{align}\label{vevdt}
\langle{\cal{O}}^{(\Delta=1)}\rangle_{\rm mABJ}
=\frac{k\mu_0}{2\pi}\sum_{n=0}^{\infty}n(n+1)(\ell^{(k)}_{n}-\ell'^{(k)}_{n})
+\cdots.
\end{align}

\section{Holographic Vacuum Expectation Values}
In section \ref{KK}, we have constructed the equations of motion \eqref{Hmn-eq4} for 4-dimensional graviton in which the graviton is coupled to the two scalar fields $T$ and $\Psi$.
The scalar field $\Psi$ is dual to the CPO in \eqref{CPO1}, whereas $T$ is dual to a gauge-invariant operator of conformal dimension two and does not play a role in the holographic calculation of the $vevs$ of ${\cal O}^{(\Delta=1)}$.
Thus, let us neglect the scalar field $T$ and consider the action for 4-dimensional graviton and the scalar field $\Psi$,
\begin{align}\label{4dact}
S= \frac1{16\pi G_4}\int d^4 x\sqrt{-g} \left(\hat R - 2\Lambda\right) - \frac{A_{\Psi}}{2}\int d^4 x\sqrt{-g}\Big(\partial_\mu \Psi\partial^\mu \Psi+ M_{\Psi}^2 \Psi^{2}\Big).
\end{align}
This action reproduces the equation of motion in \eqref{Hmn-eq4}, if the multiplicative factor is set to $\sqrt{8\pi G_4 A_{\Psi}} = \frac1{48}$.
By the field redefinition $\tilde\Psi = \sqrt{16\pi G_4 A_{\Psi}}\,\Psi$, the action in \eqref{4dact} takes canonical form,
\begin{align}\label{4dact-Res}
S= \frac{N^2}{3\sqrt{2\pi^2\lambda} L^2}\int d^4 x\sqrt{-g} \left[\hat R - 2\Lambda -\frac12\Big(\partial_\mu\tilde\Psi\partial^\mu \tilde\Psi+ M_{\Psi}^2 \tilde\Psi^{2}\Big)\right],
\end{align}
where the 4-dimensional gravitational coupling $G_{4}$ is expressed as
\begin{align}\label{G4}
\frac1{16\pi G_4} = \frac{N^2}{3\sqrt{2\pi^2\lambda} L^2}
\end{align}
with the 't Hooft coupling constant $\lambda = N/k$ in the ABJM theory~\cite{Aharony:2008ug}.
The solution for the rescaled field can be read from the asymptotic expansion of the LLM geometries in \eqref{HmnTPsi},
\begin{align}\label{tPsi1}
\tilde \Psi &= -\frac1{\sqrt{2}}\beta_3\mu_0 z + {\cal O}(\mu_0^3).
\end{align}

In odd dimensions, the Gubser, Klebanov, Polyakov-Witten (GKP-W) relation~\cite{Gubser:1998bc,Witten:1998qj} states that the $vev$ of a gauge invariant operator with conformal dimension $\Delta$ is obtained via the holographic renormalization procedure~\cite{Balasubramanian:1999re,deHaro:2000vlm,Skenderis:2000in,Bianchi:2001kw,Henningson:1998gx,deBoer:1999tgo,Kraus:1999di,Bianchi:2001de,Martelli:2002sp,Skenderis:2002wp} as
\begin{align}\label{vCPOD}
\langle {\cal O}^{(\Delta)} \rangle_{{\rm HR}} = \frac{N^2}{3\sqrt{2\pi^2\lambda}} \left(2\Delta - d\right) \tilde\psi_{\Delta},
\end{align}
where $\tilde\psi_\Delta$ is the coefficient of $z^\Delta$ term in the asymptotic expansion of the scalar field $\tilde\Psi$.
Plugging \eqref{beta3} and \eqref{tPsi1} into \eqref{vCPOD} results in the holographic $vev$ of the CPO of conformal dimension one~\cite{Jang:2016tbk} 
\begin{align}\label{vCPO12}
\langle {\cal O}^{(\Delta=1)} \rangle_{{\rm HR}} =\frac{N^2\beta_3\mu_0}{6\pi\sqrt{\lambda}}=\frac{N^{3/2}\sqrt{k}A_3\mu_0}{2\pi A_2^{3/2}}.
\end{align}

Subsequently, we employ the gauge/gravity duality dictionary and discuss the effect of the asymptotic discrete torsion $l$ on the {\it{vevs}}.

\subsection{Comparison of gravity and field theory results}
In order to figure out the role of the asymptotic discrete torsion $l$ in the holographic calculation of the $vev$, we compare the field theory result in \eqref{vevdt} to the holographic result in \eqref{vCPO12}.
Since it is not clear how to evaluate the infinite summations in \eqref{AV} and \eqref{vevdt}, it is difficult to make the comparison with full generality.
We explicitly show our results by using the example of LLM geometries represented by a rectangular Young diagram in figure 2 and then discuss the general picture. 
\begin{figure}[!h]
\centerline{\epsfig{figure=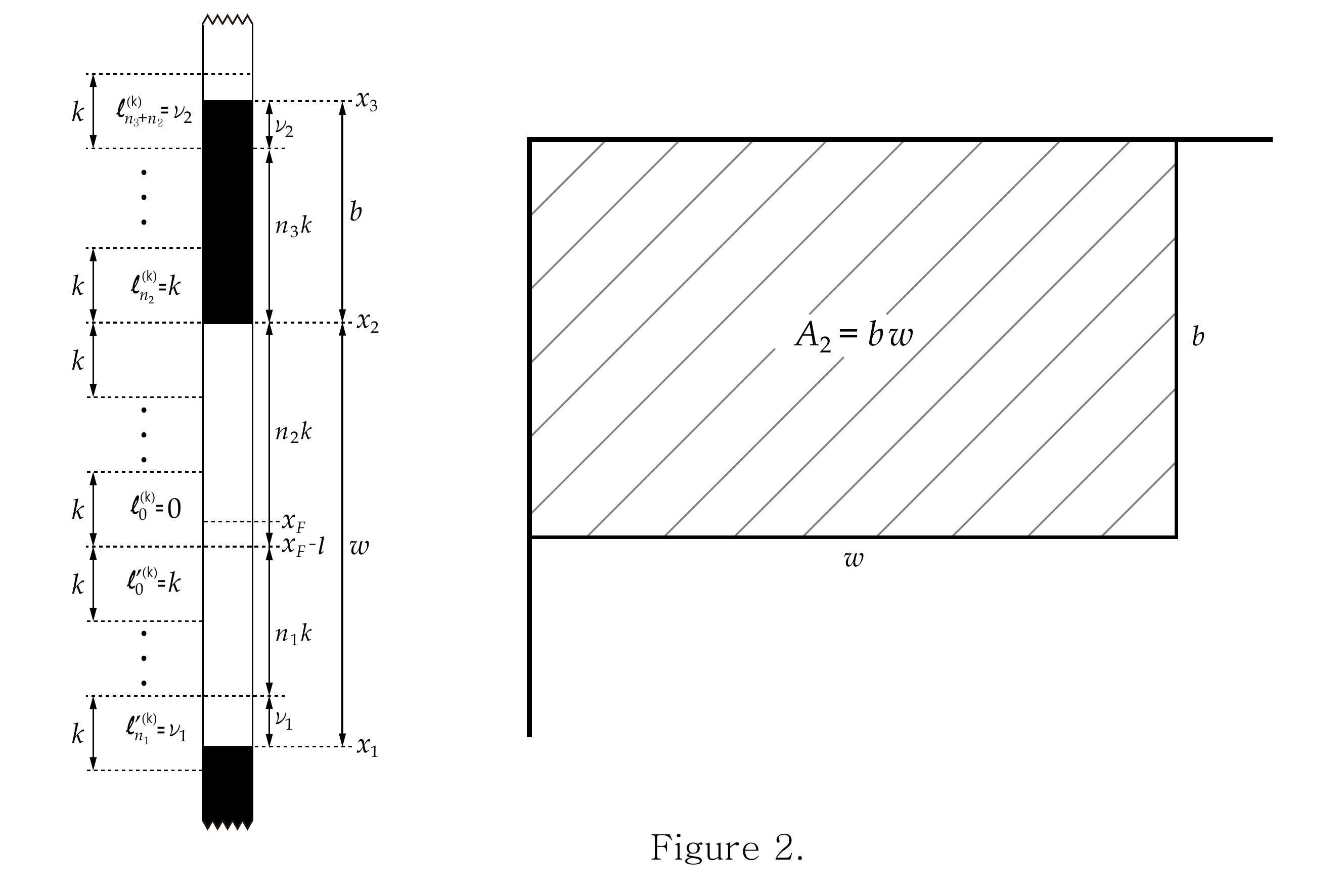,
height=100mm}}
\end{figure}

From the droplet picture in figure 2, we read the values for the discrete torsions in the following table,
\begin{align}\label{lns}
\begin{aligned}
\hline
&
\{\ell^{(k)}_{n},\ell'^{(k)}_{n}\}
&&&
\textrm{range of }n
\\
\hline
\hline
&
\ell^{(k)}_{n}=0,
&&&
0\leq n\leq n_2-1
\\
\hline
&
\ell^{(k)}_{n}=k,
&&&
n_2\leq n\leq n_2+n_3-1
\\
\hline
&
\ell^{(k)}_{n}=\nu_2,
&&&
n=n_2+ n_3
\\
\hline
&
\ell^{(k)}_{n}=0,
&&&
n\geq n_2+n_3+1
\\
\hline
&
\ell'^{(k)}_{n}=k,
&&&
 0\leq n\leq n_1-1
 \\
\hline
&
\ell'^{(k)}_{n}=\nu_1,
&&&
n=n_1
\\
\hline
&
\ell'^{(k)}_{n}=0,
&&&
n\geq n_1+1
\\
\hline
\end{aligned}
\end{align}
where $\nu_1$ and $\nu_2$ are non-negative integers smaller than $k$.
Using the relations $w=(n_1+n_2)k+\nu_1$ and $b=n_3k+\nu_2$, together with the level-matching condition in \eqref{lmc}, we obtain
\begin{align}
n_1=\frac{b-l-\nu_1}k,\qquad n_2=\frac{w-b+l}k,\qquad n_3=\frac{b-\nu_2}k,
\end{align}
where $b$ and $w$ are the lengths of the finite black and white regions, respectively.
For later convenience, we set $w=\delta\,b$ without loss of generality for some finite and positive $\delta$ and rewrite $n_2$ as
\begin{align}\label{n2}
n_2=\frac{(\delta-1)b+l}k.
\end{align}

In~\cite{Hyun:2013sf}, it was argued that the LLM geometries are highly curved if they are represented by Young diagrams that contain some short sides.
A way to avoid such highly curved geometries is to assume that both $b$ and $w$ are of the order of $\sqrt{N}$.
Actually, the relation between $(b,w)$ and $N$ is obtained by plugging the values of the discrete torsions \eqref{lns} into \eqref{NM2},
\begin{align}\label{b}
b=\sqrt{\frac{\nu_1^2+\nu_2^2-l^2+(2N-\nu_1-\nu_2+l)k}{2\delta}}.
\end{align}

The $vev$ of the CPO of conformal dimension $\Delta=1$ is  evaluated in the mABJ theory by plugging the discrete torsions \eqref{lns} into \eqref{vevdt}, accompanied by the expansion in large $N$ limit,
\begin{align}\label{FTvev}
\langle {\cal O}^{(\Delta=1)}\rangle_{\rm mABJ}=&\,\frac{\mu_0N^{3/2}}{2\pi}\bigg(\sqrt{\frac k\delta}(\delta-1)+\frac {2l}{\sqrt N}+\frac{3l(k-l)(\delta-1)}{4\sqrt{\delta k}N}\nn\\
&\quad\quad\quad\,\,\,
-\frac1{4\sqrt{\delta k}N}\Big[\nu_1(k-\nu_1)(1+3\delta)-\nu_2(k-\nu_2)(3+\delta)\Big]+{\cal O}\big(N^{-3/2}\big)\bigg)\nn\\
&+\cdots,
\end{align}
where we truncated the expansion at $\frac{1}{N}$ order for simplicity.
Similarly, the holographic $vev$ is obtained by plugging the discrete torsions \eqref{lns} into \eqref{vCPO12}.
However, now when we expand in the large $N$ limit, the subleading terms are cancelled and only the leading term survive\footnote{Following~\cite{Jang:2017gwd}, we have also tested the case of CPO of conformal dimension $\Delta=2$ and we found the holographic $vev$ is still $N$ exact and is given by
\begin{align}\label{delta2}
\langle {\cal O}^{(\Delta=2)} \rangle_{{\rm HR}} =-\frac{\mu_0^2N^{3/2}}{6\pi} \bigg[\frac{\sqrt{k}}{\delta}(\delta+1)^2\bigg].
\end{align} }
\begin{align}\label{HRvev}
\langle {\cal O}^{(\Delta=1)}\rangle_{\rm HR}&=\frac{\mu_0N^{3/2}}{2\pi}\bigg[\sqrt{\frac k\delta}(\delta-1)\bigg].
\end{align}
Comparing the field theory and the holographic $vevs$, we confirm an agreement of the two results in the large $N$ limit.

We have made similar comparisons for LLM geometries with more complex Young diagrams and the behavior is the same as that of the rectangular Young diagram.
For instance, the results for the LLM geometries represented by a Young diagram with four long sides are the following.
When we have set the lengths of the sides as $(w_1=\delta_1b,~b_1=\delta_2b,~ w_2=\delta_3b,~ b_2=b)$, $b$ is determined by \eqref{NM2},
\begin{align}
b=\sqrt{\frac{\nu_1^2+\nu_2^2+\nu_3^2-l^2+k(2 N-\nu_1-\nu_2-\nu_3 +l)}{2(\delta_1+\delta_1\delta_2+\delta_3)}},
\end{align}
where $\nu_i$'s are non-negative integers smaller than $k$. Then we obtain 
\begin{align}\label{FTvev2}
\langle {\cal O}^{(\Delta=1)}\rangle_{\rm mABJ}
=
\frac{\mu_0N^{3/2}}{2\pi}\Bigg(
&
\frac{\sqrt{k}\Big[\delta_1 (\delta_2+1) (\delta_1-\delta_2-1)+(2 \delta_1+\delta_3-1)\delta_3\Big]}{(\delta_1+\delta_1\delta_2+\delta_3)^{3/2}}+\frac{2 l} {\sqrt{N}}
\nn\\
&
+\frac{3l(k-l)\Big[\delta_1 (\delta_2+1) (\delta_1-\delta_2-1)+(2 \delta_1+\delta_3-1) \delta_3\Big]}{4\sqrt{k} N (\delta_1+\delta_1\delta_2+\delta_3)^{3/2}}
\nn\\
&
-\frac{\nu_1(k-\nu_1)\Big[\delta_1(\delta_2+1)(3 \delta_1+\delta_2+1) +(6 \delta_1 +4 \delta_2 +3 \delta_3+1)\delta_3\Big]}{4  \sqrt{k} N (\delta_1+\delta_1\delta_2+\delta_3)^{3/2}}
\nn\\
&+\frac{\nu_2(k-\nu_2) \Big[\delta_1(\delta_2+1)(\delta_1+3\delta_2+1) -(2 \delta_1 +3 \delta_3+1)\delta_3\Big]}{4  \sqrt{k} N (\delta_1+\delta_1\delta_2+\delta_3)^{3/2}}
\nn\\
&+\frac{\nu_3(k-\nu_3)\Big[\delta_1(\delta_2+1) (\delta_1+3 \delta_2+3)+ (4 \delta_1 \delta_2+2 \delta_1+\delta_3+3)\delta_3\Big]}{4\sqrt{k} N (\delta_1+\delta_1\delta_2+\delta_3)^{3/2}}
\nn\\
&+{\cal O}\left(N^{-3/2}\right)\Bigg)+\cdots,
\end{align}
and
\begin{align}\label{HRvev2}
&\langle {\cal O}^{(\Delta=1)}\rangle_{\rm HR}=\frac{\mu_0N^{3/2}}{2\pi}\frac{\sqrt{k}  \Big[\delta_1 (\delta_2+1) (\delta_1-\delta_2-1)+(2 \delta_1+\delta_3-1) \delta_3\Big]}{(\delta_1+\delta_1\delta_2+\delta_3)^{3/2}}.
\end{align}
Again, the results \eqref{FTvev2} and \eqref{HRvev2} agree in the large $N$ limit.

Absence of the subleading terms in \eqref{HRvev}, \eqref{delta2}, and \eqref{HRvev2} shows that $\langle{\cal O}^{(\Delta)}\rangle_{\rm HR}$ is an exact result in 11-dimensional tree-level supergravity.  In~\cite{Jang:2016tbk}, we have verified that there is an exact agreement between the classical field theory result  and the holographic result for the special case $k=1$. In such special case, our result here replicates those results, because all $\nu_i$'s, which are non-negative integers smaller than $k$, are zero.

\subsection{$\frac{1}{\sqrt{N}}$ correction from the discrete torsion}
In order to reveal the significance of $l$ in the holographic results, it is necessary to express \eqref{HRvev} in terms of the parameters of the theory, that can be achieved by eliminating $\delta$ from \eqref{HRvev} by using \eqref{n2} and \eqref{b}.
Then we obtain
\begin{align}\label{HRvev3}
\langle {\cal O}^{(\Delta=1)}\rangle_{\rm HR}&=\frac{\mu_0N^{3/2}}{2\pi}\bigg[\frac{\sqrt{2k}(n_2k-l)}{\sqrt{\nu_1^2+\nu_2^2-l^2+(2N-\nu_1-\nu_2+l)k}}\bigg].
\end{align}
If $n_2 k\ll \sqrt{N}$, the leading term is of the same order as the $l$-dependent term, and thus both are vanishing in the large $N$ limit.
This case corresponds to the LLM geometries represented by Young diagrams that are almost a square and the $vev$ is vanishing in the case.
On the other hand, if the Young diagrams are considered far from being a square, $n_2 k=\alpha\sqrt{N}$ for some finite value $\alpha$ and thus the $vev$ does not vanish and becomes
\begin{align}\label{HRvev5}
\langle {\cal O}^{(\Delta=1)}\rangle_{\rm HR}=\frac{\mu_0N^{3/2}}{2\pi}\bigg[\alpha-\frac{l}{\sqrt{N}}+{\cal O}\left(\frac1{N}\right)\bigg].
\end{align}
It means the asymptotic discrete torsion $l$ affects the holographic $vev$ only in the subleading order.

\section{Conclusion}
We have investigated the duality between the 3-dimensional mABJ theory and the 11-dimensional supergravity on LLM geometries with asymptotic discrete torsion $l$.
We have calculated the $vev$ of the CPO with conformal dimension $\Delta=1$ from the classical vacuum solutions of the mABJ theory and from the asymptotic limit of the LLM geometries with ${\mathbb Z}_k$ orbifold.
In our previous works~\cite{Jang:2016tbk}, we have verified that, in the special case of $k=1$, there is an exact agreement between the result obtained from the field theory classical vacuum solutions and from the holographic methods, whereas for LLM geometries with orbifold singularity, such agreement is reached only in the large $N$ limit.

We have showed that the holographic $vev$ can be expressed by infinite summations involving the discrete torsions $\{\ell_n^{(k)},\ell_n'^{(k)}\}$, which are fixed by the details of the droplet pictures representing the LLM geometries.
The {\it vevs} for CPOs with $\Delta=1,2$ depend on the shapes of droplets except for the overall $N^{3/2}$-factor.
Relying on the examples of LLM geometries represented by some specific droplet pictures, we have shown that these infinite summations can be evaluated and the holographic $vev$ is exactly determined.
However, the $vev$  obtained from the field theory classical vacuum solutions, in addition to the leading order term which agrees with the holographic $vev$, it contains $\frac{l}{\sqrt{N}}$ corrections when $k>1$ and $l\ne 0$.
This means that there is an agreement between the field theory and the holographic results in the large $N$ limit.
We have also tested our holographic calculations for the CPO of conformal dimension $\Delta=2$.

The holographic methods involve the asymptotic expansion of the LLM geometries.
The asymptotic limit of the LLM geometries with the ${\mathbb Z}_k$ orbifold is the AdS$_4\times S^7/{\mathbb Z}_k $ geometry with the radius $\left(L=(32\pi^2 A_2)^{1/6}l_{\rm P}\right)$ of the $S^7$ determined by the details of the droplet picture of the original LLM geometry.
In the $k=1$ case the radius depends only on $N$ and the asymptotic geometry does not depend on the details of the droplet picture of the LLM geometries.
Therefore, in the absence of the orbifolding, all LLM geometries with the same value of $N$ have the same asymptotic geometry, whereas such degeneracy is removed  for the LLM geometries with the orbifold singularity.

\section*{Acknowledgements}
This work was supported by the National Research Foundation of Korea(NRF) grant with grant number NRF-2018R1D1A1B07048061 (D.J.), NRF-2019R1F1A1056815 (Y.K.), NRF-2017R1D1A1A09000951 (O.K.), and NRF-2017R1D1A1B03032523 (D.T.).

\end{document}